\documentstyle[multicol,aps,pre,floats]{revtex}

\newcommand{\be}[1]{\begin{equation} \label{#1}}
\newcommand{\ee}{\end{equation}}
\newcommand{\th}{\theta}
\newcommand{\et}{\eta}
\newcommand{\tht}{\tilde{\theta}}
\newcommand{\ett}{\tilde{\eta}}
\newcommand{\ep}{\epsilon}
\newcommand{\al}{\alpha}

\input{psfig}
\draft

\title{Self-replication and splitting of domain patterns in
  reaction-diffusion systems with fast inhibitor}
\author{C. B. Muratov}
\address{Department of Physics, Boston University, Boston,
  Massachusetts, 02215}
\date{\today}
\begin{document}
\maketitle

\begin{abstract}
  An asymptotic equation of motion for the pattern interface in the
  domain-forming reaction-diffusion systems is derived. The free
  boundary problem is reduced to the universal equation of non-local
  contour dynamics in two dimensions in the parameter region where a
  pattern is not far from the points of the transverse instabilities of
  its walls. The contour dynamics is studied numerically for the
  reaction-diffusion system of the FitzHugh-Nagumo type. It is shown
  that in the asymptotic limit the transverse instability of the
  localized domains leads to their splitting and formation of the
  multidomain pattern rather than fingering and formation of the
  labyrinthine pattern.
\end{abstract}
\bibliographystyle{prsty}

\begin{multicols}{2}

\section{introduction}

Pattern formation is a remarkable phenomenon typical for many physical,
chemical, and biological systems inside and outside of thermal
equilibrium
\cite{nicolis,cross93,ko:book,ko:ufn89,ko:ufn90,vasiliev,murray,seul95}.
As a rule, these are extremely complicated phenomena. However, in many
cases pattern formation may be explained on the basis of systems of
reaction-diffusion equations of the activator-inhibitor type. The
systems of this kind include electron-hole and gas plasma, semiconductor
and superconductor structures, systems with uniformly generated
combustion material, chemical reactions with auto-catalysis and
cross-catalysis, models of morphogenesis and population dynamics (see,
for example, \cite{cross93,ko:book,ko:ufn89,ko:ufn90,vasiliev,murray}
and references therein).  The simplest example of such a system is a
pair of reaction-diffusion equations:
\be{1} 
\tau_\th {\partial \th \over \partial t} = l^2 \Delta
\th - q(\th, \et, A), 
\ee 
\be{2} \tau_\et {\partial \et \over \partial
  t} = L^2 \Delta \et - Q(\th, \et, A), 
\ee
where $\th$ is the activator, $\et$ is the inhibitor, $l$ and $L$ are
the characteristic length scales, and $\tau_\th$ and $\tau_\et$ are the
characteristic time scales of the activator and the inhibitor,
respectively, $q$ and $Q$ are certain non-linear functions, and $A$ is
the bifurcation parameter. 

Kerner and Osipov showed that the properties of patterns and pattern
formation scenarios in the systems described by Eqs. (\ref{1}) and
(\ref{2}) are chiefly determined by the parameters $\ep \equiv l/L$ and
$\al \equiv \tau_\th / \tau_\et$ and the shape of the nullcline of the
equation for the activator \cite{ko:book,ko:ufn89,ko:ufn90}. In many
cases this nullcline is N-shaped (Fig. 1). 
\begin{figure*}[htb]
\centerline{\psfig{figure=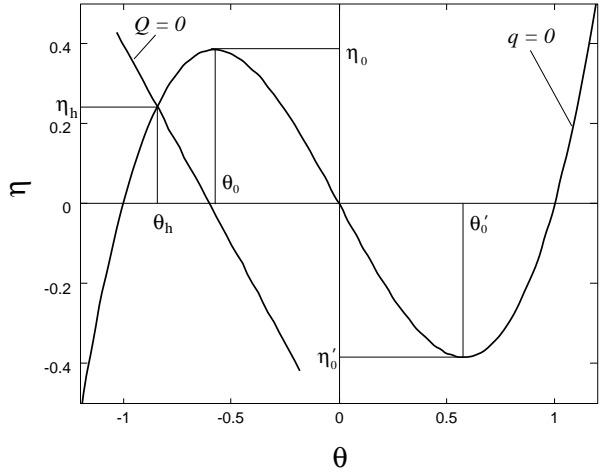,width=8cm}}
\narrowtext
\caption{Nullclines of Eqs.~(1) and (2) with $q = \th^3 - \th - \et$ and
  $Q = \th + \et - A$.}
\label{null}
\end{figure*}
\noindent
In such N systems static domain patterns may form when $\ep \ll 1$
\cite{ko:book,ko:ufn89,ko:ufn90,mo1:pre96}. These patterns are
essentially the domains of high and low values of the activator
separated by the narrow walls whose width is of order $l$. Recent
experiments and numerical simulations have revealed a lot of new pattern
formation scenarios in these systems such as growth of fingers, tip
splitting, spot replication, and formation of labyrinthine patterns
\cite{epstein95,lee:pre95,hagberg:prl94,hagberg:chaos94,%
  goldstein96,mo2:pre96}.  These effects are associated with the fact
that at certain parameters the patterns become unstable with respect to
the transverse perturbations. Muratov and Osipov developed a general
asymptotic theory of such instabilities in an arbitrary N system
described by Eqs.  (\ref{1}) and (\ref{2}) \cite{mo1:pre96}. They have
shown that the instabilities are determined by the motion of the pattern
walls.  Goldstein, Muraki, and Petrich derived an equation of motion for
a simple reaction-diffusion system of FitzHugh-Nagumo type in the limit
of fast inhibitor and small activator-inhibitor coupling and drew
similar conclusions about the transverse instabilities of the domain
patterns \cite{goldstein96}. The numerical simulations of concrete
reaction-diffusion systems in the limit of fast inhibitor ($\al \gg
\ep$) showed that the instabilities typically lead to the formation of
labyrinthine patterns. Spot replication was observed only in the case of
slow inhibitor ($\al \lesssim \ep$) \cite{hagberg:chaos94,mo2:pre96}.

Numerical solution of Eqs. (\ref{1}) and (\ref{2}) for a concrete model
shows that for the same values of the parameters there exist
qualitatively different types of stable static solutions. These are the
solutions in the form of the labyrinthine patterns [Fig. 2(a)] and the
multidomain patterns [Fig. 2(b)]. The labyrinthine pattern forms as a
result of the instability of a single domain which is taken as an
initial condition in Fig. 2(a). In the simulation of Fig. 2(b) the
initial condition was taken in the form of a random arrangement of
domains. As a result, a metastable multidomain pattern forms in the
system. We emphasize that the only difference between the simulations of
Fig. 2 is the initial conditions, all parameters characterizing the
system itself are identical in both cases.

The aim of this paper is to show that in the asymptotic limit $\ep
\rightarrow 0$ the transverse instabilities of the pattern walls will
always lead to splitting of domains and formation of the multidomain
patterns in the limit of fast inhibitor. To show this we will reduce the
partial differential equations problem in an arbitrary $d$-dimensional N
system to the free boundary problem in the limit $\ep \rightarrow 0$ for
arbitrary $\al$. We will further reduce this problem near the
instability points of the domain patterns to the problem of non-local
contour dynamics for $\al \gg \ep$ and obtain a universal equation of
motion for the interface in two dimensions. This equation will be
studied numerically for a concrete model.

\section{free boundary problem}

In this section it is convenient to measure the lengths in the units of
$l$ and time in the units of $\tau_\th$, respectively. Then
Eqs. (\ref{1}) and (\ref{2}) become:
\be{act}
{\partial \th \over \partial t} = \Delta \th - q(\th, \et, A),
\ee
\be{inh}
\al^{-1} {\partial \et \over \partial t} = \ep^{-2} \Delta \et - Q(\th,
\et, A).
\ee

Ohta, Mimura, and Kobayashi developed an approach which allowed them to
reduce equations similar to Eqs. (\ref{act}) and (\ref{inh}) to the
problem of the interface dynamics in the case of slow inhibitor ($\al
\lesssim \ep$) in the limit $\ep \rightarrow 0$ \cite{ohta89}. Here we
will use their approach to derive the equations of the interface dynamics
for Eqs. (\ref{act}) and (\ref{inh}) and show that these equations are
essentially the same in the case of fast and slow inhibitor.

Let us introduce the local coordinate system in the vicinity of the
interfaces of the pattern (Fig. \ref{scheme}). For a point $x$ let
$\rho$ be the distance from the point $x$ to the interface, and the
$(d-1)$-dimensional coordinate $\xi$ the projection of $x$ on the
submanifold $\rho = const$. $\rho$ is assumed to be positive if the
point is in the region of high activator values and negative otherwise.
The distribution of the activator varies on the length scale 1 near the
interface. Since the characteristic length of the variation of the
inhibitor is $\ep^{-1} \gg 1$, $\et$ can be considered constant in the
direction perpendicular to the interface: $\et = \et_i(\xi)$. From the
general considerations follows that the curvature of the interface
$K(\xi)$ has to be small \cite{ko:book,mo1:pre96}.
\end{multicols}
\widetext
\begin{figure*}[htb]
\centerline{\psfig{figure=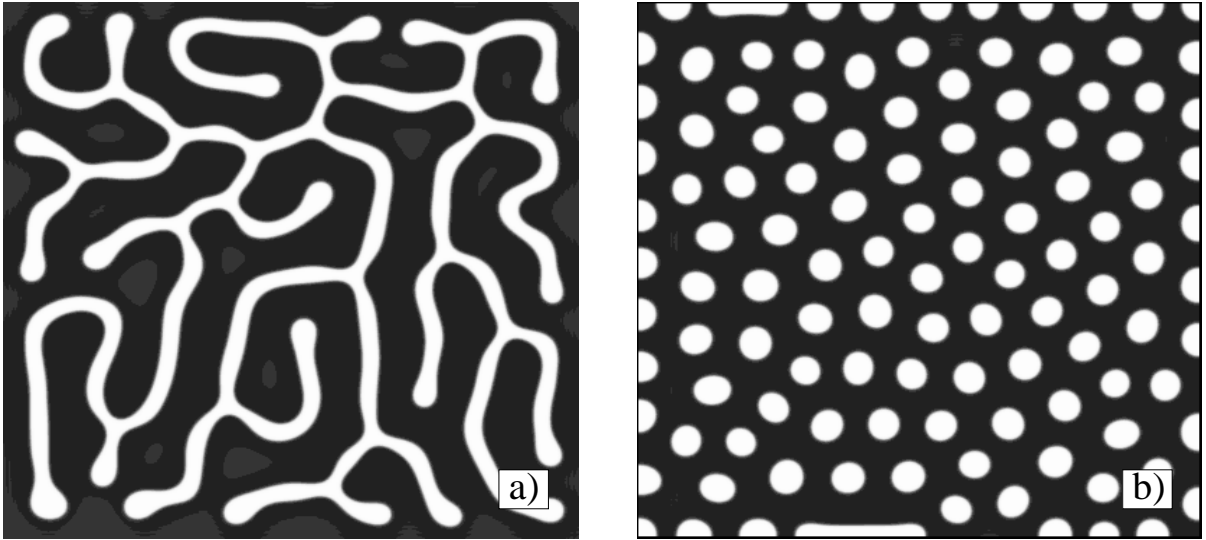,width=16cm}}
\vspace{0.5cm}
\caption{Two types of extended domain patterns: labyrinthine pattern (a) and
  multidomain pattern (b). Results of the numerical simulations of
  Eqs. (1) and (2) with $q = \th^3 - \th - \et$ and $Q = \th + \et -
  A$. The parameters used are: $\ep = 0.05, \al = 0.2, A = -0.4$. The
  system size is $20L \times 20L$. The time of the simulation is $500
  \tau_\et$. For explanations of the initial conditions see the text. 
  }
\label{fig2}
\end{figure*}
\begin{multicols}{2}
\noindent
Therefore, to the first power in $\ep$ and $K$ we may write Eq.
(\ref{act}) as follows:
\be{act:sh}
{\partial \th \over \partial t} = {\partial^2 \th \over \partial \rho^2}
- K(\xi) {\partial \th \over \partial \rho} - q(\th, \et_i(\xi)).
\ee
The sign of $K$ is such that it is positive if the interface is convex
into the low activator region. Notice that $\xi$ enters only as a
parameter in this equation. For times $t \gg 1$ any solution of
Eq. (\ref{act:sh}) connecting high and low values of the activator will
become close to the solution in the form of a front propagating with the
velocity $v$ in the direction perpendicular to the interface. For any
given point $\xi$ one can then write 
\be{auto}
-v(\xi) {d \th \over d z} = {d^2 \th \over dz^2} + K(\xi) {d \th \over
  dz} - q(\th, \et_i(\xi)),
\ee
where we introduced the self-similar variable $z = -\rho - v t$. The
velocity is positive when it points out of the high activator value
domain. The boundary conditions for this equation are:
\be{bc}
\th(+\infty) = \th_{i1}(\xi), ~~~\th(-\infty) = \th_{i3}(\xi),
\ee
where $\th_{i1,3}$ are the minimal and the maximal roots of the equation
\be{eta:i}
q(\th_{i1,3}(\xi), \et_i(\xi)) = 0
\ee
for any given $\xi$. The third solution $\th_{i2}(\xi)$ of
Eq. (\ref{eta:i}) which lies between $\th_{i1}$ and $\th_{i3}$ may be
used to fix the position of the interface relative to $\th(z)$ by
requiring that $\th = \th_{i2}(\xi)$ on the interface. Thus, $v(\xi)$
represents the normal velocity of the pattern interface at a point
$\xi$. 

The velocity $v$ can be found by solving Eq. (\ref{auto}) with the
abovementioned boundary conditions. The consistency condition which is
obtained from Eq. (\ref{auto}) by multiplying it by $d \th / dz$ and
integrating over $z$ gives the following expression for the velocity in
terms of $\th(z)$: 
\be{v}
v = -K - {\int_{\th_{i1}}^{\th_{i3}} q(\th, \et_i) d \th \over
  \int_{-\infty}^{+\infty} \left( {d \th \over dz} \right)^2 dz}.
\ee
From this equation follows that the front velocity $v$ is of order 1. 
Notice that Eq. (\ref{auto}) is essentially an equation of motion of a
particle in the potential 
\be{U}
U_\th = - \int q(\th, \et_i) d \th
\ee
with the friction proportional to $v + K$ with time $z$
\cite{ko:book,ko:ufn89,ko:ufn90}. In N systems $U_\th$ is a double hump
potential (for those values of $\et_i$ for which it has only a single
hump Eq. (\ref{auto}) has no solutions), so Eq. (\ref{auto}) has a
unique solution which satisfies the boundary conditions in Eq.
(\ref{bc}) for the given values of $\et_i$ and $K$, which corresponds to
the trajectory going from the top of one hump to the top of the other.
Thus, the velocity of the interface is a single-valued function of the
curvature and the value of $\et$ at the interface.

\begin{figure*}[htb]
\centerline{\psfig{figure=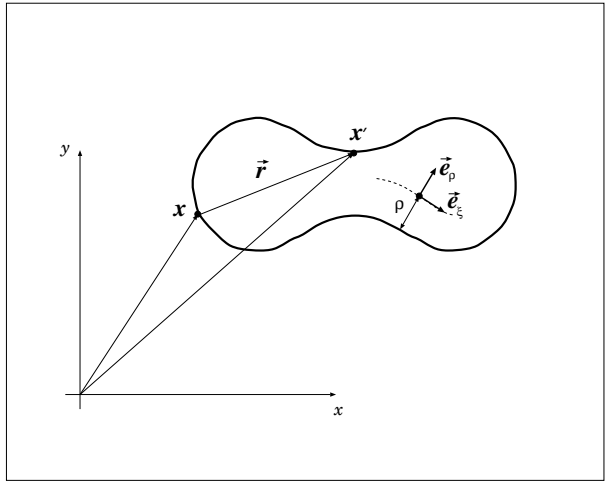,width=8cm}}
\narrowtext
\hspace{0.5cm}
\caption{The schematics of the system's geometry (two dimensions). The
  thick solid line shows the interface. $\vec{e}_\rho$ and $\vec{e}_\xi$
  are the local orthogonal basis of the curvilinear coordinate system
  $\rho, \xi$; $\rho$ indicates the distance from a given point to the
  interface, the dashed line indicates the surface $\rho = const$. }
\label{scheme}
\end{figure*}
\noindent

Far from the interfaces the activator varies on the characteristic
length $\ep^{-1}$, so the Laplacian in Eq. (\ref{act}) in that region
can be dropped and Eq. (\ref{act}) reads:
\be{sm:t}
{\partial \th \over \partial t} = - q(\th, \et).
\ee
So, the relation between $\th$ and $\et$ far from the interfaces is
local in space. Since $\et$ varies on the length scales of order
$\ep^{-1}$ or $K^{-1}$ and the velocity of the front is of order 1, the
characteristic time of the variation of the interface velocity is also
$\ep^{-1} \gg 1$ or $K^{-1} \gg 1$, what justifies the use of Eq.
(\ref{auto}) for finding the distribution of the activator near the
interfaces. The characteristic time of variation of the inhibitor field
caused by the interface motion will also be $\ep^{-1}$ or $K^{-1}$ in
the case $\al \gg \ep$ (fast inhibitor), or $\al^{-1}$, which is even
greater than the previous in the case $\al \lesssim \ep$ (slow
inhibitor), so the time derivative in Eq. (\ref{sm:t}) is less or of
order $\ep$ and, therefore, can be neglected. Then, the values of $\th$
and $\et$ far from the interfaces are simply related by the equation of
local coupling:
\be{lc}
q(\th, \et) = 0.
\ee
In view of Eq. (\ref{lc}), the inhibitor must satisfy the equation: 
\be{sm}
\al^{-1} {\partial \et \over \partial t} = \ep^{-2} \Delta \et -
Q(\th(\et), \et),
\ee
with the following boundary conditions at the interfaces (written in
terms of the local curvilinear coordinates $\rho$ and $\xi$): 
\be{bc:i}
\et(+0, \xi) = \et(-0, \xi) = \et_i(\xi). 
\ee
The dependence $\th(\et)$ in Eq. (\ref{sm}) is the solution of
Eq. (\ref{lc}). This dependence is multi-valued, so for the regions of
high and low activator values one should take the branches connected
with $\th_{i3}$ and $\th_{i1}$, respectively. Thus, we recovered the
result obtained by Ohta, Mimura, and Kobayashi in Ref. \cite{ohta89} in
the case of fast inhibitor as well. The term in the left-hand side of
Eq. (\ref{sm}) is of order $\ep / \al$ which can be set equal to zero in
the case of fast inhibitor. 

Equation (\ref{sm}) is essentially nonlinear since it involves the
inverse function of essentially nonlinear function $q(\th, \et)$, even
if the Eq. (\ref{inh}) were linear. Moreover, the right-hand side of
Eq. (\ref{sm}) becomes singular at the points $\th_0$ and $\th_0'$ (see
Fig. \ref{null}) where $q'_\th(\th_0, \et_0) = 0$ and $q'_\th(\th_0',
\et_0') = 0$, respectively, since the function $\et(\th)$ becomes
singular at these points. When the distribution of $\et$ at a point
$x_0$ reaches the values of $\et_0$ or $\et_0'$, a sudden down-jump or
up-jump (local breakdown), occurs at that point, respectively 
\cite{ko:book,ko:ufn89,ko:ufn90,mo1:pre96,mo2:pre96}. In terms of
the free boundary problem this corresponds to the instantaneous creation
of a new interface at the point $x_0$, which will start to evolve
according to the equation of the interface motion. This is an important
ingredient of the free boundary problem describing the dynamics of the
pattern interface that should not be left out in solving this
problem. Notice that local breakdown is responsible for the domain
splitting in the one-dimensional N systems
\cite{ko:book,ko:ufn89,ko:ufn90}. 

Equations (\ref{auto}) -- (\ref{eta:i}) together with
Eqs. (\ref{lc}) -- (\ref{bc:i}) and the rule for dealing with the
singularities of Eq. (\ref{sm}) discussed in the previous paragraph
are the closed set of equations which defines completely the dynamics of
any domain pattern in the system described by Eqs. (\ref{act}) and
(\ref{inh}) in the limit $\ep \rightarrow 0$ both in the case of slow
and fast inhibitor. These equations were obtained with the accuracy to
$\ep \ll 1$ and $K \ll 1$. Note that the presence of the curvature term
in the equation of motion for the interface suggests only that the
interface has to be sufficiently smooth, more exactly, it has to be
smooth on the length scales of order 1. This, of course, does not mean
that the interfaces are not allowed to intersect, fuse, or loose their 
connectivity in the course of their evolution. Self-replicating domains, 
which will be studied in the subsequent sections, are a good example of
the latter. Also, a domain may disappear if it shrinks too much. 

\section{non-local contour dynamics}

As was shown in Ref. \cite{mo1:pre96}, when $\al \gg \ep$ static
domain patterns may be stable only when their characteristic size is 
much smaller than the characteristic length of the inhibitor variation
because of the stabilizing action of the interaction between the 
walls of the pattern. Typically, a pattern destabilizes when the
characteristic distance between its walls (in the units of the previous
Section) becomes of order $\ep^{-2/3}$ \cite{mo1:pre96}. In this
situation the velocity of the non-stationary pattern interface must be
much smaller than one. Also, the value of the inhibitor must be close to
the value of $\et_s$ at which the velocity of the pattern wall is equal
to zero in one dimension. According to Eq. (\ref{v}) with $K = 0$, the
value of $\et_s$ must satisfy
\be{eta:s}
\int_{\th_{s1}}^{\th_{s3}} q(\th, \et_s, A) d \th = 0,
~~~~~q(\th_{s1,3}, \et_s, A) = 0,
\ee 
where $\th_{s1}$ and $\th_{s3}$ are some constants, which generally
depend on $A$. This situation takes place when the bifurcation parameter
$A$ is close to the values of $A_b$ or $A_b'$ where the characteristic
distance between the walls of the domains with high or low values of the
activator becomes zero in the limit $\ep \rightarrow 0$. 

For definiteness let us consider a single domain of high values of the
activator. Then the value of the activator inside the domain will be
close to $\th_{s3}$, and to $\th_{s1}$ outside. Let us introduce the new
variables:
\be{tilde}
\tht = \th - \th_{s1}, ~~~~\ett = \et - \et_s
\ee
Equations (\ref{lc}) and (\ref{sm}) can then be linearized. In the case
of fast inhibitor we will have:
\be{lc:l}
q'_\th(\th_{s1}, \et_s) \tht + q'_\et(\th_{s1}, \et_s) \ett = 0,
\ee
\be{sm:l}
0 = \ep^{-2} \Delta \ett - C \ett - a I(x) - Q(\th_{s1}, \et_s),
\ee
where 
\be{C}
C = Q'_\et(\th_{s1}, \et_s) - {Q'_\th(\th_{s1}, \et_s) q'_\et(\th_{s1},
  \et_s) \over q'_\th(\th_{s1}, \et_s)},
\ee
\be{a}
a = Q(\th_{s3}, \et_s) - Q(\th_{s1}, \et_s),
\ee
and $I(x)$ is the indicator function which is equal to 1 if $x$ is
inside the domain and 0 outside. The value of $Q(\th_{s1}, \et_s)$ is
small for $A$ close to $A_b$. Note that in writing Eq. (\ref{sm:l})
we neglected the piecewise-constant potential inside the
domains which is present upon the linearization in the general case
since one has to linearize Eq. (\ref{lc}) inside and outside of the
domain separately. However, this is justified when the domain size is
much smaller than the characteristic length of the inhibitor variation
since the potential then is only a perturbation and can be neglected in
the zeroth order.  

Since the velocity of the front is small, it can also be linearized in
$\ett$. This, however, extremely simplifies the problem since one no
longer has to solve the non-linear eigenvalue problem in
Eq. (\ref{auto}). Indeed, in the linear approximation in $\ett$ we can
replace $\th(z)$ in the denominator of Eq. (\ref{v}) by the solution in
the form of the static one-dimensional front and expand the numerator in
$\ett$, so we immediately obtain for the velocity at a point $\xi$ on
the interface:
\be{v:l}
v(\xi) = - K(\xi) - \ett_i(\xi) Z^{-1} \int_{\th_{s1}}^{\th_{s3}}
q'_\et(\th, \et_s) d 
\th, 
\ee
and
\be{Z}
Z = \int_{\th_{s1}}^{\th_{s3}} \sqrt{2 (U_\th(\th_{s1}, \et_{s}) -
  U_\th(\th, \et_s) )} d \th,
\ee
where $U_\th$ is defined in Eq. (\ref{U}). 

The distribution of $\ett(x)$ at each moment in time and, therefore, the
values of $\ett$ on the interface, which determine the interface
velocity, can be found by solving Eq. (\ref{sm:l}) by means of the 
Green's function: 
\be{ett}
\ett(x) = -{Q(\th_{s1}, \et_s) \over C} - a \int G(x - x') I(x') d^d x'. 
\ee
Specifically, in the infinite two-dimensional system
\be{G2}
G(x - x') = \frac{\ep^2}{2 \pi} K_0( | x - x'| \ep \sqrt{C} ),
\ee
where $K_0$ is the modified Bessel function. Following the idea of
Ref. \cite{goldstein96}, let us transform the integral in
Eq. (\ref{ett}) in two dimensions into the contour integral along the
interface. Using the defining equation for the Green's function and the
Gauss theorem, we find: 
\begin{eqnarray} \label{cont}
\ett(x) = && - {Q(\th_{s1},  \et_s) \over C} 
\nonumber \\
&&
+ \frac{a \ep}{2 \pi \sqrt{C}}
\oint \biggl( K_1( \ep r \sqrt{C} ) 
- \frac{1}{\ep r \sqrt{C} } \biggr) {\vec{r} \times d \vec{r}
\over r}
, 
\end{eqnarray}
where $\vec{r}$ is the vector from the point $x$ to $x'$, where the
point $x'$ lies on the interface (see Fig. \ref{scheme}), and the
integration is over the interface, $K_1$ is the modified Bessel
function. In writing Eq.  (\ref{cont}) it was taken into account that
the surface integral obtained from Eq. (\ref{G2}) can as well be written
in terms of the vector product involving the tangent vector $d \vec{r}$.
The interface has to be oriented counterclockwise in order to get the
sign right. If there is more than one domain in the system, one has to
add up the contributions of each domain given by the integral in Eq.
(\ref{cont}).

For a single domain of the size of order $\ep^{-2/3}$ one can expand the
Bessel function in Eq. (\ref{cont}), so we obtain the following equation
of motion for the interface:
\begin{eqnarray} \label{spot}
v(\xi) = -K(\xi) && - {B Q(\th_{s1}, \et_s) \over a C Z}
\nonumber \\
&& + \frac{B \ep^2}{4 \pi Z} \oint (\ln (0.54 \ep
\sqrt{C}) + \ln r ) \vec{r} \times d \vec{r},
\end{eqnarray}
where
\be{B}
B = -(Q(\th_{s3}, \et_s) - Q(\th_{s1}, \et_s) )
\int_{\th_{s1}}^{\th_{s3}} q'_\et( \th, \et_s) d \th,
\ee
and the point $x$ is now on the interface. If we introduce the
renormalized coordinates, and time and introduce the new control
parameter $\tilde{A}$: 
\be{renorm}
\tilde{x} = x \ep^{2/3}, ~~\tilde{t} = t \ep^{4/3},~~\tilde{A} = - { B 
  Q(\th_{s1}, \et_s) \ep^{-2/3} \over a C Z},
\ee
we will eliminate the $\ep$-dependence in Eq. (\ref{spot}) (except for
the weak logarithmic dependence). Thus, we obtained the asymptotic
equation of motion for a localized domain with the characteristic size
much smaller than the characteristic size of the inhibitor variation.
This equation was derived with the accuracy to $\ep^{2/3}$. Note that at
this point all information about the specific nonlinearities of the
system is contained only in a few numerical constants: $B$, $C$ and $Z$.
These constants are all positive for the reaction-diffusion systems of
the activator-inhibitor type \cite{mo1:pre96}. Furthermore, they can be
incorporated into the rescaled $\tilde{\ep} = \ep Z C^{3/2} / B$,
provided that time and coordinate are also suitably rescaled. So, the
dynamics of the localized domains in any two-dimensional N system not
far from $A_b$ depends only on $\tilde{A}$ and $\tilde{\ep}$, the last
dependence being logarythmically weak.

Equation (\ref{spot}) is also good for the description of several
interacting domains if the distance between them is not much greater
than $\ep^{-2/3}$. In order to study the interaction of domains
separated by the distances of order $\ep^{-1}$ one has to use
Eqs. (\ref{v:l}) and (\ref{cont}).  

Equation (\ref{spot}) contains a large logarithm which multiplies the
integral $\oint \vec{r} \times d \vec{r}$ which up to a coefficient is
the area of the domain. This means that with the logarithmic accuracy
the area of the evolving domain has to be conserved. Also, the non-local
term in Eq. (\ref{spot}) is an increasing function of $r$, so the
parts of the interface which are close to each other attract, and the
parts which are far from each other repel. Then the instability which
appears for a radially-symmetric domain when its radius reaches certain
value \cite{mo1:pre96} cannot lead to the growth of a labyrinthine
pattern, but rather will result in the domain splitting. The domains
will then go apart and the process of splitting will repeat, until the
systems becomes filled with the multidomain pattern. This mechanism of
the instability development is qualitatively different from the one
discussed in Ref. \cite{goldstein96}.

\section{self-replicating domains in a concrete reaction-diffusion
  system} 
In this Section we will apply the results obtained above and show that
the proposed mechanism of the instability development indeed takes place
in a concrete reaction-diffusion system. To do this we will use the
system which is described by Eqs. (\ref{act}) and (\ref{inh}) with
\be{q}
q = \th^3 - \th - \et,
\ee
\be{Q}
Q = \th + \et - A.
\ee
The nullclines of this system for $A = -0.6$ are shown in Fig.~1. 

This system is particularly convenient because the dependences of its
characteristic parameters on $A$ are especially simple. Also, the
non-linear eigenvalue problem in Eq. (\ref{auto}) in this system admits
exact solution. Simple calculation shows that:
\be{s}
\et_s = 0, ~~~\th_{s1} = -1, ~~~\th_{s3} = 1,
\ee
and the coefficients involved in Eq. (\ref{spot}) are:
\be{coeff}
a = 2, ~~~B = 4, ~~~C = \frac{3}{2}, ~~~Z = \frac{2 \sqrt{2}}{3}.
\ee
The value of $Q(\th_{s1}, \et_s)$ becomes zero at $A = A_b$ with $A_b =
-1$. Also, the homogeneous state of the system $\th_h = -|A|^{1/3},
\et_h = -|A|^{1/3} (1 - |A|^{2/3})$ is stable for $A < A_c = - 1/ 3
\sqrt{3}$.  

The solution of Eq. (\ref{auto}) can be sought in the form $\th(z)
= a \tanh bz + c$, where $a$, $b$, and $c$ are constants. As a result,
the dependence of the front velocity on $\et_i$ (with $K = 0$) is
implicitly given by the following equation: 
\be{v3}
v - \frac{2}{9} v^3 = \frac{3}{\sqrt{2}} \et_i.
\ee
The velocity $v$ satisfies $|v| < \sqrt{3/2}$, the maximum value being
achieved at $\et = \et_0$, where $\et_0 = 2/3 \sqrt{3}$. Thus, for all
values of $\et$ for which Eq. (\ref{sm}) is not singular the dependence
$v(\et_i)$ is single-valued.

Another special feature of this system is the fact that in it the 
piecewise-constant potential mentioned in Sec. III is identically
zero, so Eqs. (\ref{v:l}) and (\ref{cont}) also describe the dynamics of
any complex pattern with the characteristic domain size much smaller
than $\ep^{-1}$ for any values of $A$. These include the complex
patterns forming in the late stages of the development of the transverse
instabilities \cite{mo1:pre96,mo2:pre96}. In fact, the analog of Eq.
(\ref{cont}), non-local both in space and time, may be obtained also in
the case of arbitrary $\al$. This equation, together with Eq.
(\ref{v:l}), could be used to study the pulsations (breathing) of
complex domain patterns in this system.

The direct numerical simulations of Eqs. (\ref{act}) and (\ref{inh}) is
a formidable task. The main difficulty here is the fact that for small
$\ep$ there are two very different length scales, so the simulations
require enormous amounts of computational power. Recently, it became
possible to perform extensive numerical simulations of the system under
consideration using massive parallelization on a supercomputer
\cite{mo2:pre96}. The authors were able to simulate the system of
sufficiently large sizes with $\ep \simeq 0.05$. The values of $\ep
\simeq 0.01$ are already very hard to simulate even on the very fast
computer. 

\end{multicols}
\widetext
\begin{figure*}[htb]
\centerline{\psfig{figure=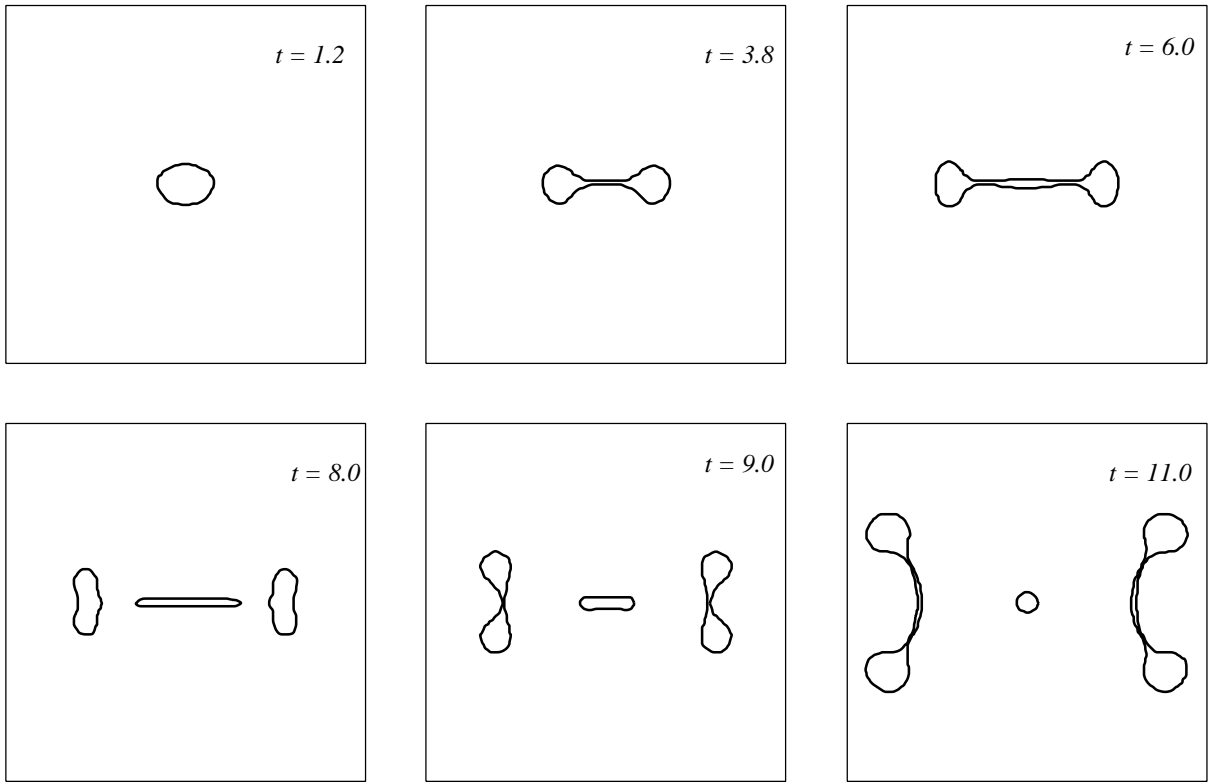,width=16cm}}
\vspace{0.5cm}
\caption{
  Destabilization of the circular domain. Results of the numerical
  simulation of Eq. (26) for the considered model with $\tilde{A} = 20$
  and $\ep = 10^{-4}$. The box indicates the size of $30 \times 30$. The
  length and time are measured in the rescaled units given by Eq. (28).
  }
\label{fig4}
\end{figure*}

\widetext
\begin{figure*}[htb]
\centerline{\psfig{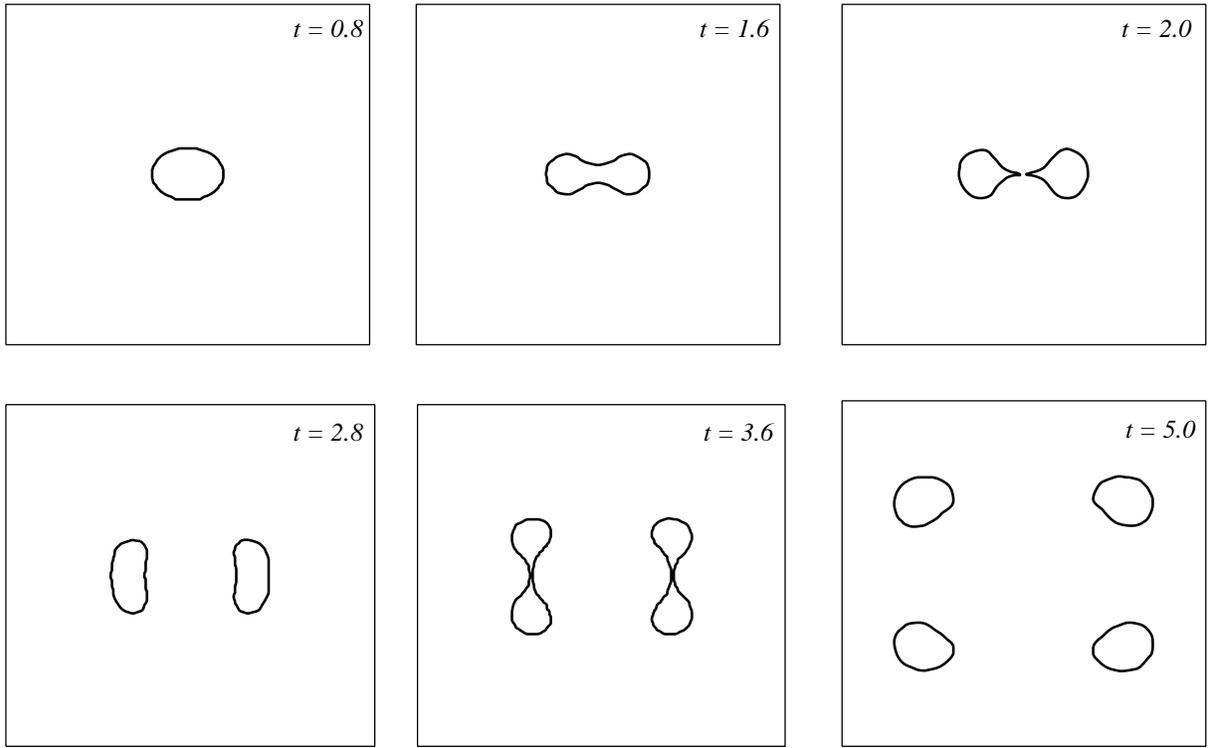}}
\vspace{0.5cm}
\caption{
  Self-replication of the domains. Results of the numerical simulation
  of Eq. (26) for the considered model with $\tilde{A} = 30$ and $\ep =
  10^{-4}$. The box indicates the size of $30 \times 30$. The length and
  time are measured in the rescaled units given by Eq. (28).  }
\label{fig5}
\end{figure*}

\begin{multicols}{2}
\noindent
The main result of these simulations relevant to the present discussion
is that as a result of the development of the transverse instability a
localized domain transforms into a labyrinthine pattern, which is all
connected if $A$ is not far from $A_b$. This effect takes place at $\ep
> 0.01$. For those values of $\ep$ domain splitting and self-replication
was observed only when $\al \lesssim \ep$ \cite{mo2:pre96}. We performed
numerical simulations of this system for $\ep \leq 0.01$ and saw that a
destabilizing localized domain actually splits into two.  However, the
simulation becomes excessively long at this point, so it is impossible
to see if the forming domains will split in turn.

Equation (\ref{spot}), from the other hand, allows to study the
interfacial dynamics for arbitrarily small $\ep$. We performed numerical
simulations of Eq. (\ref{spot}) for the values of the parameters when
a single localized domain becomes unstable with respect to transverse
perturbations of its walls.  

Figure \ref{fig4} shows the evolution of the almost circular domain when
its radius is close to the critical radius at which the domain loses
stability with respect to the $m = 2$ mode. 
Initially the domain elongates, but at some point the distance between
the walls becomes so small that the domain splits into the three
disconnected pieces. The resulting domains continue to grow until the
larger domains split again into seven (not shown in the Figure), and the
process goes on. Notice that in order for this process to take place,
the value of $\ep$ has to be very small. As is seen from the numerical
simulations of Eqs.  (\ref{act}) and (\ref{inh}), when $\ep$ is not very
small the domain can elongate and transform into a stripe without
splitting. This effect also takes place in the reaction-diffusion
systems with the weak activator-inhibitor coupling \cite{goldstein96}.

Figure \ref{fig5} shows the results of another simulation when the
system is further away from the instability point. There the initial
stage of the domain growth is similar to the one shown in
Fig. \ref{fig4}. However, the walls in the center come closer to each
other than in the previous case, so at some point the domain splits into
two. The resulting domains continue to grow and split in turn. This
process is essentially self-replication of the domains, as a result of
which the system will become filled with the multidomain pattern. 

We emphasize that unless the system is very close to the point $A =
A_b$, the localized domains will always be unstable and, once excited,
will transform into a multidomain pattern via self-replication even in
the case of fast inhibitor, if $\ep$ is small enough. This is a
completely universal result independent of any property of the system.
In particular, this conclusion does not depend on whether the system is
monostable or bistable.  As was noted in the previous Section, the only
nontrivial system-dependence is contained in the logarithmic term, and,
therefore, this dependence is weak.

The results obtained for the system under consideration do not change in
a wide range of $\ep$. When $\ep \gtrsim 0.01$, Eq. (\ref{spot}) seizes
to be a good approximation for Eqs. (\ref{v:l}) and (\ref{cont}). For
extremely small $\ep$ ($\ep \lesssim 10^{-9}$) there exist a narrow
region of the values of $\tilde{A}$ close to the critical value at which
the disk transforms into a stable domain in the form of a dumbbell. When
the value of $\tilde{A}$ is increased, the domain self-replication will
occur even for such small values of $\ep$.

\section{conclusion}

Thus, we have shown that in the case of fast inhibitor the transverse
instability of the localized domains in the reaction-diffusion systems
with N-shaped nullcline for the activator will always lead to domain
splitting and formation of multidomain pattern rather than the formation
of labyrinthine patterns, if $\ep$ is sufficiently small. 

This effect was in fact observed in a quasi two-dimensional experiment
with the current filaments forming in n-GaAs in the process of avalanche
breakdown \cite{mayer88}. There the radially-symmetric current filaments
destabilized and split as the current in the sample increased and the
radius of a filament grew. At some critical value of current a filament
split into two and the filaments that formed split in turn until their
radii become sufficiently small. Non-symmetrically distorted or
elongated filaments were not observed. 

Hagberg and Meron suggested that splitting of domains is the consequence
of the parity-breaking front transitions [nonequilibrium Ising-Bloch
(NIB) transitions] associated with the variations of the curvature of
the domain walls \cite{hagberg:chaos94}. However, their arguments may
apply only to the bistable reaction-diffusion systems in which the
inhibitor is slow. We showed here that the splitting of domains in fact
occurs in the systems with the {\em fast} inhibitor and is determined by
the non-local interaction between different portions of the domain
interface, regardless of whether the system is monostable or bistable,
so as a rule the NIB transitions should not be responsible for domain
splitting. It is clear that in the systems with the slow inhibitor
domain splitting will occur even easier since the inhibitor will not be
able to react on the motion of the walls of the domains locally not only
in space, but also in time.  This is also confirmed by the direct
numerical simulations of Eqs.  (\ref{1}) and (\ref{2}) \cite{mo2:pre96}.

Notice that the condition $\ep \ll 1$ itself is the necessary condition
for the existence of the static domain patterns in the considered systems
\cite{ko:book,ko:ufn89,ko:ufn90}, so, in fact, the transition from a
localized domain to the multidomain pattern consisting of disconnected
localized domains filling up the space must be the major mechanism of
the transverse instability development. Multidomain patterns were indeed
observed in the chemical systems \cite{epstein95} and in the
high-frequency gas discharge \cite{ammelt93}. Nevertheless, numerical
simulations and experimental observations also show that for small but
finite $\ep$ localized domains may transform into extended labyrinthine
patterns \cite{epstein95,lee:pre95,hagberg:prl94,goldstein96,mo2:pre96}.
This does not seem to be the consequence of the finite width of the
interface, but rather a peculiarity of the inhibitor dynamics. This
effect can actually be controlled by varying the strength of interaction
between the activator and the inhibitor [the constant $B$ in Eq.
(\ref{spot})].  Goldstein, Muraki, and Petrich analyzed the equation
similar to Eqs.  (\ref{v:l}) and (\ref{cont}) in the case $B \sim \ep$
in the limit $\ep \rightarrow 0$ and found that the transverse
instability leads to the formation of the connected labyrinthine
pattern.  This suggests that by changing $\ep$ or the coupling strength
$B$ one could control whether the multidomain pattern or the
labyrinthine pattern will form as a result of the domain instability.

It is important to note that the free boundary problem obtained from
Eqs. (\ref{1}) and (\ref{2}) in the limit $\ep \ll 1$ contains
considerably less information about the nonlinearities of the system
than the initial partial differential equations problem. Moreover,
according to Eq. (\ref{spot}), the behavior of any localized pattern not
far from the points $A_b$ or $A_b'$ is universal in the sense that the
dynamics of the interface can be described by only a few renormalized
parameters. This universality was discussed earlier in the context of
the instabilities of the domain patterns \cite{mo1:pre96}. This suggests
that the free boundary formulation of the pattern dynamics might be a
more advantageous starting point for dealing with the problems of domain
pattern formation rather than the formulation in terms of
reaction-diffusion equations.

The phenomenology of pattern formation similar to the one discussed in
the present article is also observed in various equilibrium systems with
competing interactions (see, for example, \cite{seul95} and references
therein). Notice that the reaction-diffusion system described by Eqs.
(\ref{1}) and (\ref{2}) in the limit $\tau_\et \rightarrow 0$ describes
the kinetics of a system with competing interactions, if $q = f(\th) -
\et$ and $Q = \et + B \th$, where $B$ is a constant and $f(\th)$ is some
cubic-like function \cite{ohta86,ohta90,goldstein96}. These equations
describe, for example, the kinetics of microphase separation of block
copolymers \cite{ohta86,ohta95}. The universality of the results
obtained above suggests that self-replication of domains must be a
common feature of the systems with competing interactions in the case of
repulsive long-range interactions of Coulombic type and strong
separation of length scales.

The author is grateful to V. V. Osipov for valuable discussions. 

\bibliography{../main}

\begin{thebibliography}{10}

\bibitem{nicolis}
G. Nicolis and I. Prigogine, {\em Self-organization in Non-Equilibrium Systems}
  (Wiley Interscience, New York, 1977).

\bibitem{cross93}
M. Cross and P.~C. Hohenberg, Rev. Mod. Phys. {\bf 65},  851  (1993).

\bibitem{ko:book}
B.~S. Kerner and V.~V. Osipov, {\em Autosolitons: a New Approach to Problems of
  Self-Organization and Turbulence} (Kluwer, Dordrecht, 1994).

\bibitem{ko:ufn89}
B.~S. Kerner and V.~V. Osipov, Sov. Phys. -- Uspekhi {\bf 32},  101  (1989).

\bibitem{ko:ufn90}
B.~S. Kerner and V.~V. Osipov, Sov. Phys. -- Uspekhi {\bf 33},  679  (1990).

\bibitem{vasiliev}
V.~A. Vasiliev, Y.~M. Romanovskii, D.~S. Chernavskii, and V.~G. Yakhno, {\em
  Autowave Processes in Kinetic Systems: Spatial and Temporal Self-Organization
  in Physics, Chemistry, Biology, and Medicine} (VEB Deutscher Verlag der
  Wissenschaften, Berlin, 1987).

\bibitem{murray}
J.~D. Murray, {\em Mathematical Biology} (Springer-Verlag, Berlin, 1989).

\bibitem{seul95}
M. Seul and D. Andelman, Science {\bf 267},  476  (1995).

\bibitem{mo1:pre96}
C.~B. Muratov and V.~V. Osipov, Phys. Rev. E {\bf 53},  3101  (1996).

\bibitem{epstein95}
I.~R. Epstein and I. Lengyel, Physica D {\bf 84},  1  (1995).

\bibitem{lee:pre95}
K. Lee and H. Swinney, Phys. Rev. E {\bf 51},  1899  (1995).

\bibitem{hagberg:prl94}
A. Hagberg and E. Meron, Phys. Rev. Lett. {\bf 72},  2494  (1994).

\bibitem{hagberg:chaos94}
A. Hagberg and E. Meron, Chaos {\bf 4},  477  (1994).

\bibitem{goldstein96}
R.~E. Goldstein, D.~J. Muraki, and D.~M. Petrich, Phys. Rev. E {\bf 53},  3933
  (1996).

\bibitem{mo2:pre96}
C.~B. Muratov and V.~V. Osipov,   (in preparation).

\bibitem{ohta89}
T. Ohta, M. Mimura, and R. Kobayashi, Physica D {\bf 34},  115  (1989).

\bibitem{mayer88}
K.~M. Mayer, J. Parisi, and R.~P. Huebner, Z. Phys. B -- Cond. Matter {\bf 71},
   171  (1988).

\bibitem{ammelt93}
E. Ammelt, D. Schweng, and H.-G. Purwins, Phys. Lett. A {\bf 179},  348
  (1993).

\bibitem{ohta86}
T. Ohta and K. Kawasaki, Macromolecules {\bf 19},  2621  (1986).

\bibitem{ohta90}
T. Ohta, A. Ito, and A. Tetsuka, Phys. Rev. A {\bf 42},  3225  (1990).

\bibitem{ohta95}
T. Ohta and A. Ito, Phys. Rev. E {\bf 52},  5250  (1995).

\end{thebibliography}

\end{multicols}
\end{document}